\documentclass[aps,prb,twocolumn,amsmath,amssymb,floatfix]{revtex4}

\usepackage{graphicx}

\newcommand{\corr}[1]{\ensuremath{\left\langle #1 \right\rangle}}

\begin{document}


\title{Exact results on the Kondo-lattice magnetic polaron}

\author{S.~Henning}
 \email{henning@physik.hu-berlin.de}
\author{P.~Herrmann}
\author{W.~Nolting}
\affiliation{Lehrstuhl Festk\"orpertheorie, Institut f\"ur Physik, Humboldt-Universit\"at zu Berlin, Newtonstrasse 15, 12489 Berlin, Germany} 

\date{\today}

\begin{abstract}
In this work we revise the theory of one electron in a ferromagnetically saturated local moment
system interacting via a Kondo-like exchange interaction. The complete
eigenstates for the finite lattice are derived. It is then shown, that parts of
these states lose their norm in the limit of an infinite lattice. The correct
(scattering) eigenstates are calculated in this limit. The time-dependent
Schr\"odinger equation is solved for arbitrary initial conditions and the
connection to the down-electron Green's function and the scattering states is
worked out. A detailed analysis of the down-electron decay dynamics is given. 
\end{abstract}

\pacs{Valid PACS appear here}
\maketitle
\section{introduction}
The Kondo-lattice model (KLM) has found widespread application as a basic model for the
description of itinerant carriers interacting with local magnetic moments formed
by inner shells of the constituting atoms. It was successfully applied in the
theoretical description of the europium chalcogenides \cite{Methfessel1968,Nolting1979,Wachter1979}, 
aspects of the physics of manganites \cite{Salamon2001,Stier2007,Stier2008,Held2000}
and the diluted magnetic (III,Mn)V semiconductors \cite{Jungwirth2006,Stier2011}.

It has long been known that the KLM, although it forms a complicated many-body
problem not solvable in general, has a non-trivial, exactly solvable
limiting case - the (ferro-)magnetic polaron\cite{Methfessel1968,Izumov1971,Richmond1970}. 
This magnetic polaron is set up by adding one electron into the otherwise empty
conduction band and a ferromagnetically saturated background.

Shastry and Mattis\cite{Shastry1981} gave a detailed discussion of the
spectral properties of the polaron in terms of the retarded one-electron Green's
function (GF).  Berciu and Sawatzky \cite{Berciu2009} have extended the GF
solution to describe complex lattices and longer ranged exchange interactions.
Knowledge of the GF allows for the direct calculation of the bare line shape of
a photo emission experiment\cite{Nolting2009} via the spectral density (SD).
This strength of the GF approach comes at the expense of getting only indirect information
about the underlying eigenstates of the one-electron quantum system.

Therefore other approaches have solved Schr\"odinger's equation directly to get this state
information\cite{Izumov1971,Methfessel1968,Nolting1979a,Shastry1981}. 
Sigrist~et.~al.\cite{Sigrist1991} gave a rigorous proof, that the
ground state for a system with antiferromagnetic exchange coupling is of
incomplete ferromagnetic order. We found these derivations to be incomplete
in one way or another and there is, to our knowledge, 
no exhaustive derivation of the eigenstates
for the finite system of $N$ lattice sites in the literature until now.

In the limit $N\to\infty$ the free dispersion of the electron
becomes a continuous function of the wavevector. It was van Hove who pointed out
in his seminal papers \cite{Hove1955,Hove1957} that in case of a continuous (free) spectrum
the interaction part of the Hamiltonian can lead to persistent effects not
amenable to scattering theory. This self-energy effects can be of dissipative
nature (finite lifetime of quasi-particles) and/or cloud effects (formation of a new quasi-particle with
infinite lifetime). Both effects are present in the problem of the magnetic
polaron.

The phenomenon of decaying states has attracted the interest of many physicists.
Far from being comprehensive here we just want to mention the works, that have
inspired us to the present investigation. 

Nakanishi \cite{Nakanishi1958} proposed an extended quantum theory, that contains complex eigenvalues of the
Hamiltonian which can be associated with an unstable particle. To this aim he
constructed the analytic continuation of the propagator for that particle and
found poles in the continuation. By introducing a complex distribution he was
able to construct a new eigenstate of the Hamiltonian with complex eigenvalue
(equal to the pole position) leading to an exponential decay of the particle in
time. Deviations from an exponential decay law at long times and the importance of the van Hove
singularity on the lower bound of the spectrum was discussed by 
H\"ohler\cite{Hoehler1958} and Khalfin\cite{Khalfin1958}. The short-time
deviations and the resulting quantum Zeno effect where derived in
\cite{Misra1977,Chiu1977}. Sudarshan~et.~al. have extended the ideas of
Nakanishi\cite{Sudarshan1978} and gave also a more formal derivation in a rigged
Hilbert space formalism\cite{Parravicini1980,Sudarshan1993}.

We will use the methods developed in the above mentioned works as a pragmatic
device to get a deeper understanding of the decay dynamics of a down-electron
in the problem of the magnetic polaron.

The paper is organized as follows.\\
In section \ref{sec:model} we explain the model Hamiltonian and give the
parameters used throughout this work.
Section \ref{sec:GF} summarizes the results obtained with the GF approach.
The spectral density of up/down-electrons is discussed in detail.
The complete eigenstates for the finite lattice are derived in section
\ref{sec:eigenstates}.
Section \ref{sec:thermodynlim} is devoted to the problems arising in the
limit of an infinite lattice. New (scattering) eigenstates are constructed and the
time-dependent Schr\"odinger equation is solved for arbitrary initial
conditions. The connection of certain initial conditions with the electronic GF is
given. 
In section \ref{sec:dod} a detailed analysis of the dynamics of the quantum
system for two different initial conditions is given. The connection to the
scattering states and the limitations of scattering theory are discussed.
Finally we give a summary and draw conclusions in section \ref{sec:sumcon}.
\section{model}
\label{sec:model}
Throughout the paper we are concerned with the s-f-like Hamiltonian:
\begin{equation}
H = \sum_{ij\sigma}T_{ij}c_{i\sigma}^+ c_{j\sigma}
-\frac{J}{2}\sum_{i\sigma}\left(z_{\sigma}S^{z}_{i}c_{i\sigma}^+ c_{i\sigma} 
+S^{\sigma}_{i}c_{i-\sigma}^+ c_{i\sigma}\right)
\label{eq:hamiltonian}
\end{equation}
where $\sigma=\pm$, $z_{\pm}=\pm1$ and $S^{\pm}_i = S^{x}_i \pm iS^{y}_i$, 
describing free electrons hopping trough
the lattice and undergoing a local (contact) interaction with immobile local
moments formed by the inner shells of the underlying atoms. The
$c_{i\sigma}^+$ ($c_{i\sigma}$) denote the creation (annihilation) operator of an
electron with spin $\sigma$ at lattice site $\mathbf{R}_i$.\\
The above Hamiltonian constitutes a highly involved many-body problem not solvable
for the general case. It has long been known, however, that there exists a 
non-trivial solution in the limiting case of one electron in the otherwise empty conduction band
moving in a lattice of fully ferromagnetically  ordered background spins
($T=0K$).\\ 
The Hamiltonian (\ref{eq:hamiltonian}) does not distinguish energetically between different
configurations of the localized spins in case of an empty band. One could
enforce a fully aligned ground state by introducing an additional
direct exchange term of Heisenberg type between the localized spins as was
done in some works \cite{Shastry1981}. This additional term would lead to a true magnon
dispersion of the localized moments (which is flat in our case) and to
energy corrections of the eigenstates where a magnon is present. Since these
energy corrections are small (typical magnon energies are 2-3 orders of
magnitude smaller than the electronic energies) and the additional term would
complicate the already involved calculations we omit this term and choose the
fully aligned state out of the highly degenerate ground-state manifold
``by hand''. One has to keep in mind, however, that the dynamics of the localized
spins are mediated only by the electron.\\
If not stated otherwise we will use the following model-parameters for the
concrete evaluation of our theory.
We choose a magnetic moment $S=\frac{7}{2}$ for the local spin system which
reflects the magnetic moment of the prototypical rare-earth compounds EuO and
EuS. The underlying
lattice will be the three-dimensional (3D) simple cubic (SC) lattice and the
hopping $T_{ij}$ is chosen to give the tight-binding dispersion:
\begin{equation}
\epsilon_{\mathbf{k}} = -\frac{W}{6}\left(\cos(kx)+\cos(ky)+\cos(kz)\right)
\label{eq:dispersion}
\end{equation}
with bandwidth $W=1$.
\section{one-electron green's function}
\label{sec:GF}
\begin{figure}
    \includegraphics[width=0.95\linewidth]{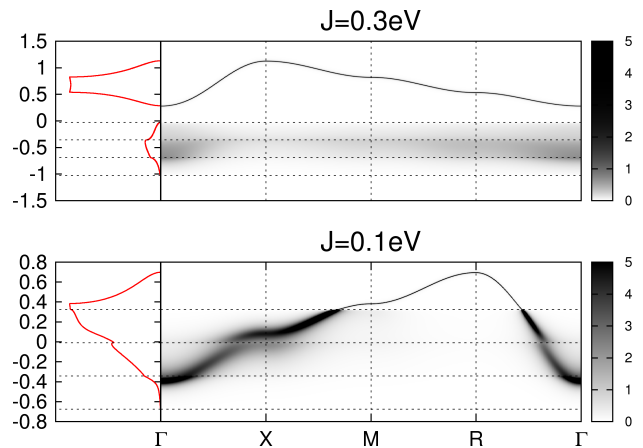}
    \caption{\label{fig1}(color online) Down electron spectral density along lines of high symmetry in
    $\mathbf{k}$-space and quasi-particle density of states for two different
    coupling strengths $J$. Parameters: $S=3.5$, $W=1.0$ eV.}
\end{figure}
The limiting case of the ``magnetic polaron'' is usually discussed in terms of the 
retarded one-electron Green's function (GF). 
For the derivation of the latter one transforms (\ref{eq:hamiltonian})
into reciprocal $\mathbf{k}$-space and writes down the equation of motion
(EQM) of the GF. Under the above assumptions (ferromagnetic saturated local
moment system, empty band) this EQM can be simplified in case of an up-electron (parallel to
local moments) to give:
\begin{equation}
G_{\mathbf{k}\uparrow}(E) = G^{(0)}_{\mathbf{k}}(E+\frac{JS}{2})=
\frac{1}{E+i0^{+} - \epsilon_{\mathbf{k}} + \frac{JS}{2}}.
\label{eq:Green_up}
\end{equation}
An up-electron behaves essentially like a free electron with a shifted energy equal to the
mean field value due to the ferromagnetic background. It has an infinite
lifetime because it will not find a partner to flip its spin in the already
saturated local moment system. \\
The situation is quite different for a
down-electron. In the EQM appears a higher spin-flip GF (SF-GF) and one has to go
one step further in the EQM-hierarchy to get a closed set of equations.   
This results in:
\begin{equation}
G_{\mathbf{k}\downarrow}(E) = \frac{1}{E+i0^{+}-\epsilon_{\mathbf{k}}
-\Sigma_{\downarrow}(E)},
\label{eq:Green_down}
\end{equation}
with the electronic self-energy\cite{Shastry1981}: 
\begin{equation}
\Sigma_{\downarrow}(E) = \frac{JS}{2}\left(1+\frac{JG^{(0)}(E+\frac{JS}{2})}
                          {1-\frac{J}{2}G^{(0)}(E+\frac{JS}{2})}\right),
\label{eq:Self_Green_down}
\end{equation}
where $G^{(0)}(E)=\frac{1}{N}\sum_{\mathbf{k}}G^{(0)}_{\mathbf{k}}(E)$ denotes
the $\mathbf{k}$-summed free electronic GF. 
The center of gravity of the down-spectral density is given by the first spectral 
moment\footnote{A short summary of how the spectral moments are defined and how they 
can be calculated is given in appendix (\ref{app:spectral_m_down}).}:
\begin{equation}
M_{\mathbf{k}\downarrow}^{(1)} = \epsilon_{\mathbf{k}}+\frac{JS}{2},
\label{eq:centerofgrav}
\end{equation}
which is the mean-field energy of the down-electron. Despite this one expects
significant changes in the down-spectral density compared to the mean-field result
due to correlation effects. Especially possible spin-flips of the down-electron
should result in a finite lifetime of quasi-particles as is indicated by the
complex-valued self-energy. Inspecting (\ref{eq:Self_Green_down}) reveals, that
states with finite lifetime can be expected exactly in the energy range of the
up-spectrum because the self-energy has a finite imaginary part there. 
These states are commonly called the \textit{scattering states}.
In order to fulfill (\ref{eq:centerofgrav}) for the center of gravity
one then should find additional states (of infinite lifetime) above ($J>0$)
or below ($J<0$) the scattering states as real roots of the denominator of (\ref{eq:Green_down}) at
least for sufficiently large $J$. \footnote{A sufficient condition is, that the center
of gravity lies outside the scattering states, that is $|JS|>W/2$.}
In the next section we will show, that there is exactly one such additional
state called the \textit{bound state} or \textit{polaron state}. \\
To illustrate our findings we have plotted the down-spectral density and
the quasi-particle density of states (QDOS) in Fig.~\ref{fig1} for two
different values of $J$. For large enough $J$ ($J=0.3$ eV, upper figure) the 
bound states are completely separated from the scattering states and form a
polaron band. It can be shown \cite{Richmond1970,Izumov1971}, that in the limit of large $J$ ($|J| \gg W/S$) 
this polaron band has its center of gravity at $E_{c}^{p}=\frac{J}{2}(S+1)$ and
the bandwidth is reduced by a factor of $\frac{2S}{2S+1}$ compared to $W$.\\
When $J$ is smaller ($J=0.1$ eV, lower figure) parts of the polaron-band 
dip into the scattering states. There is still a visible polaron dispersion in
the scattering region, but now in form of a peak with finite linewidth implying
a finite lifetime of the quasi-particle. We will show in a later section
(\ref{sec:dod}), that 
each quasi-particle peak in the scattering states is associated with a pole
in the suitably analytically continued propagator leading to an exponential 
contribution in the decay dynamics. The polaron dispersion is strongly disturbed 
at the positions of the van Hove singularities (horizontal dashed lines) of the
scattering spectrum. The special role of the van Hove singularities as
limit/branch points for analytic continuation will also become clear in that section.
\section{The finite system - Eigenstates}
\label{sec:eigenstates}
In this section we derive the complete eigensystem for a finite lattice ($N$
lattice sites) with
periodic boundary conditions. Although parts of this derivation can be found
in the literature \cite{Methfessel1968,Shastry1981}, to our knowledge, the complete spectrum was never calculated.
Especially the appearance of pure up-electron states with one magnon emitted is
not recognized.\\
We use the following notation. The state of one electron with wavevector
$\mathbf{k}$, spin $\sigma$ and all local moments aligned (magnon vacuum) 
will be denoted by:
\begin{equation}
c_{\mathbf{k}\sigma}^+|0;0\rangle = |\mathbf{k}\sigma;0\rangle.
\label{eq:magnon_vac}
\end{equation}
An up-electron with wavevector $\mathbf{k}$ plus a magnon of wavevector
$\mathbf{q}$ is written as:
\begin{equation}
\frac{1}{\sqrt{2S}}S_{\mathbf{q}}^{-} c_{\mathbf{k}\uparrow}^+|0;0\rangle =
|\mathbf{k}\uparrow;\mathbf{q}\rangle.
\label{eq:magnon}
\end{equation}
These states span the Hilbert subspace we are interested in.
The Hamiltonian (\ref{eq:hamiltonian}) commutes with the 
z-component of the total spin operator $\hat{S}_{tot}^{z}=\sum_i (S_i^z +
\frac{1}{2}\sum_{\sigma}z_{\sigma}\hat{n}_{i\sigma})$. Therefore we can classify
the eigenstates by $S_{tot}^z$ and by their (outer) wavevector
$\mathbf{k}$ due to translational invariance. \\
In the subspace of $S_{tot}^z=NS+\frac{1}{2}$ the eigenstates are simply the
up-electron states with magnon vacuum:
\begin{equation}
H|\mathbf{k}\uparrow;0\rangle =
\left(\epsilon_{\mathbf{k}}-\frac{JS}{2}\right)|\mathbf{k}\uparrow;0\rangle.
\label{eq:ES_up}
\end{equation}
The subspace $S_{tot}^z=NS-\frac{1}{2}$ is more interesting. It will be spanned
by the states of one down electron in magnon vacuum and an up-electron plus one
magnon emitted. By using the following Ansatz for the wave-function:
\begin{equation}
|\Psi_{\mathbf{k}}^{n}\rangle = A\left\{\alpha_{\mathbf{k}}|\mathbf{k}\downarrow;0\rangle
+\sum_{\mathbf{q}}\beta_{\mathbf{k,q}}|\mathbf{k-q}\uparrow;\mathbf{q}\rangle\right\}
\label{eq:ansatz_wave}
\end{equation}
we get a system of equations for the coefficients from Schr\"odinger's equation:
\begin{eqnarray}
\label{eq:coeff_soe}
0 &=& \left(
    E-\epsilon_{\mathbf{k}}-\frac{JS}{2}
\right) 
    \alpha_{\mathbf{k}}+J\sqrt{\frac{S}{2N}}\sum_{\mathbf{q}}\beta_{\mathbf{k,q}} \\
0 &=& \left(
E-\epsilon_{\mathbf{k-q}}+\frac{JS}{2}
\right)\beta_{\mathbf{k,q}}+J\sqrt{\frac{S}{2N}}\alpha_{\mathbf{k}}-\frac{J}{2N}\sum_{\mathbf{q}}
\beta_{\mathbf{k,q}}. \nonumber
\end{eqnarray}
This are $N+1$ equations for the same number of coefficients and we expect $N+1$
eigenvalues per $\mathbf{k}$-value ($N(N+1)$ eigenvalues in total). The
characteristic polynomial of (\ref{eq:coeff_soe}) turns out to be of the simple form:
\begin{equation}
0 =\left(\prod_{i=1}^{N}E_i\right)
\left(E_0-\frac{J}{2}(E_0+JS)\frac{1}{N}\sum_{j}^{N}\frac{1}{E_j}\right)
\label{eq:characteristic_poly}
\end{equation}
with
\begin{eqnarray}
E_0 &=& E-\epsilon_{\mathbf{k}}-\frac{JS}{2}, \\
E_{n\ne0} &=& E-\epsilon_{\mathbf{k-q_n}}+\frac{JS}{2}.
\label{eq:E_i}
\end{eqnarray}
Not all $\epsilon_{\mathbf{k-q_n}}$ will be different by symmetry arguments.
Therefore we collect all equal $\epsilon_{\mathbf{k-q_n}}$ in groups, where the
numeric value of the group members is denoted
by $\epsilon_{\mathbf{k}}^{(n)}$ with the convention that
$\epsilon_{\mathbf{k}}^{(1)}=\epsilon_{\mathbf{k}}$.  
Let us assume that $N_G$ such groups with respective degree of 
degeneracy $g_n$ exist. The $\mathbf{q}$ belonging
to one group are denoted by $\mathbf{q}_l^{(n)}$ with $l\in0\cdots g_n-1$.
With the definition:
$F_n=E-\epsilon_{\mathbf{k}}^{(n)}+\frac{JS}{2}$ we can recast
(\ref{eq:characteristic_poly}) to give:
\begin{equation}
0=\left(\prod_{i=1}^{N_G}F_i^{g_i}\right)\left(
E_0-F_1\frac{J}{2N}\sum_{j=1}^{N_G}\frac{g_j}{F_j} \right).
\label{eq:charpol_sym}
\end{equation}
The first product contributes $N_G$ differing eigenvalues:
\begin{equation}
E_{\mathbf{k}}^{(n)}=\epsilon_{\mathbf{k}}^{(n)}-\frac{JS}{2}.
\label{eq:eigenval_pureup}
\end{equation}
The respective multiplicity of these eigenvalues will be $g_n-1$, since we divide by $F_j$ (for $j>1$) in the
second factor of (\ref{eq:charpol_sym}). Only $E_{\mathbf{k}}^{(1)}$ has the full multiplicity $g_1$.
For the construction of the eigenvectors we notice first, that the eigenenergies
(\ref{eq:eigenval_pureup}) are equal to the pure up-electron energies (\ref{eq:ES_up})
found in the $S_{tot}^{z}=NS+\frac{1}{2}$ sector of the Hilbert space.
It is therefore reasonable to assume the sought-after states to be up-states as well.
We try $\alpha_{\mathbf{k}}=0$ and $\beta_{\mathbf{k,q}}\ne 0$ only for 
$\mathbf{q}\in \{\mathbf{q}^{(n)}_l\}$ in our Ansatz (\ref{eq:ansatz_wave}).
From (\ref{eq:coeff_soe}) we then get the condition:
\begin{equation}
\sum_{\mathbf{q}\in\{\mathbf{q}^{(n)}_l\}}\beta_{\mathbf{k,q}} = 0.
\label{eq:cond_beta_up}
\end{equation}
This condition can be nontrivial fulfilled when the $\beta_{\mathbf{k,q}}$
are chosen to be suitable normalized powers of the primitive $g_n$th roots of unity.
With this we find for the eigenstates:
\begin{equation}
|\Psi_{\mathbf{k}}^{n,m}\rangle=\frac{1}{\sqrt{g_n}}
\sum_{l=0}^{g_n-1}\mathrm{e}^{i\frac{2\pi}{g_n}ml}
|\mathbf{k-q}_{l}^{(n)}\uparrow;\mathbf{q}_{l}^{(n)}\rangle
\label{eq:eigenstates_up}
\end{equation}
where $m\in1\cdots g_n-1$ is numerating the eigenstates of the subspace with
energy $E_{\mathbf{k}}^{(n)}$.\\
One eigenstate is missing so far, since the degeneracy of $E_{\mathbf{k}}^{(1)}$
is $g_1$. The states, which have only an up-electron (plus magnon) component are
exhausted by (\ref{eq:eigenstates_up}). By substituting $E_\mathbf{k}^{(1)}$ into
(\ref{eq:coeff_soe}) a simple calculation shows, that the last eigenstate is given
by:
\begin{equation}
|\Psi_{\mathbf{k}}^{1,0}\rangle=\frac{
	|\mathbf{k}\downarrow;0\rangle+\frac{\sqrt{2NS}}{g_1}
\sum_{l=0}^{g_1-1}|\mathbf{k-q}_{l}^{(1)}\uparrow;\mathbf{q}_{l}^{(1)}\rangle
}{\sqrt{1+\frac{2NS}{g_1}}}.
\label{eq:state_1_0}
\end{equation}
It has a finite down-electron component.\\ 
Until now we have found $1+\sum_{i=1}^{N_G}(g_i-1)=1+N-N_G$ eigenvalues and
corresponding eigenvectors from the first factor of (\ref{eq:charpol_sym}).
The remaining $N_G$ eigenvalues come from the second factor:
\begin{eqnarray}
\label{eq:ew_equ_2}
0 &=& E_0-F_1\frac{J}{2N}\sum_{j=1}^{N_G}\frac{g_j}{F_j}\\
  &=& E-\epsilon_{\mathbf{k}}-\frac{J}{2}\left( S+1+\frac{1}{N}\sum_{n=2}^{N_G}
  		\frac{g_n(\epsilon_{\mathbf{k}}^{(n)}-\epsilon_{\mathbf{k}})}
		     {E-\epsilon_{\mathbf{k}}^{(n)}+\frac{JS}{2}}\right).\nonumber
\end{eqnarray}
The function on the rhs of (\ref{eq:ew_equ_2}) has $N_G-1$ zero-crossings
between the up-state energies (\ref{eq:eigenval_pureup}) and one above ($J>0$)
or below ($J<0$) the scattering states near
$E_c\approx\epsilon_{\mathbf{k}}+\frac{J}{2}(S+1)$.
\begin{figure}
    \includegraphics[angle=-90, width=0.95\linewidth]{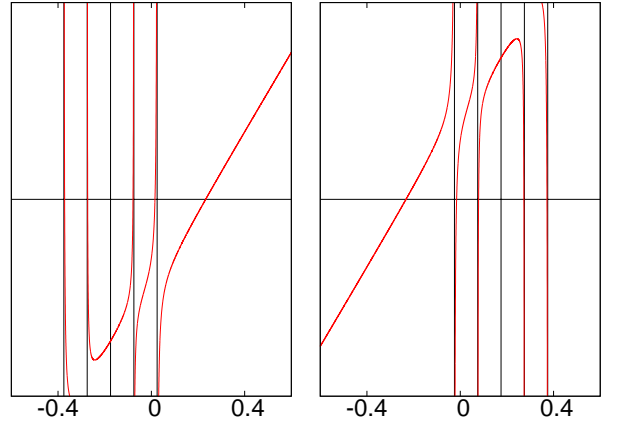}
    \caption{\label{fig2}(color online) Plot of eigenvalue equation (\ref{eq:ew_equ_2}) with
    positive (left figure) and negative (right figure) $J$.
    Parameters: $|J|=0.1$, $S=3.5$, $\epsilon_{\mathbf{k}}=0$, $\epsilon_{\mathbf{k}}^{(1)}=-0.2$, 
    $\epsilon_{\mathbf{k}}^{(2)}=-0.1$, $\epsilon_{\mathbf{k}}^{(3)}=0.1$,
    $\epsilon_{\mathbf{k}}^{(4)}=0.2$, $g_n=1$.}
\end{figure}
The situation is depicted in Fig.~\ref{fig2}. Rewriting (\ref{eq:ew_equ_2})
we get an implicit equation for the (polaron)-eigenenergies:  
\begin{equation}
E_{\mathbf{k}}^{p,(n)}=\epsilon_{\mathbf{k}}+\frac{J}{2}\left(
S+1+\Phi_{\mathbf{k}}(E_{\mathbf{k}}^{p,(n)}+\frac{JS}{2}) \right)
\label{eq:eigenval_p}
\end{equation}
with 
\begin{equation}
\Phi_{\mathbf{k}}(z)=\frac{1}{N}\sum_{\mathbf{q}}\frac{\epsilon_{\mathbf{k-q}}-\epsilon_{\mathbf{k}}}
{z-\epsilon_{\mathbf{k-q}}}.
\label{eq:phi_k}
\end{equation}
The corresponding eigenstates are again obtained from (\ref{eq:coeff_soe}) to
yield:
\begin{eqnarray}
\label{eq:eigenst_p}
|\Psi_{\mathbf{k}}^{p,(n)}\rangle & = & \mathcal{N}|\mathbf{k}\downarrow;0\rangle   \\
&-& \mathcal{N}\frac{\sum_{\mathbf{q}}\left(1+\frac{\epsilon_{\mathbf{k-q}}-\epsilon_{\mathbf{k}}}
                             {E_{\mathbf{k}}^{p,(n)}-\epsilon_{\mathbf{k-q}}+\frac{JS}{2}}
                     \right)|\mathbf{k-q}\uparrow;\mathbf{q}\rangle}
                     {\sqrt{2NS}} \nonumber
\end{eqnarray}
with the obvious normalization factor:
\begin{equation}
\mathcal{N}=\left(1+\frac{1}{2NS}\sum_{\mathbf{q}}
    \left|
        1+\frac{\epsilon_{\mathbf{k-q}}-\epsilon_{\mathbf{k}}}
             {E_{\mathbf{k}}^{p,(n)}-\epsilon_{\mathbf{k-q}}+\frac{JS}{2}}
    \right|^2\right)^{-\frac{1}{2}}.
\label{eq:norm_polstates}
\end{equation}
\section{The infinite system}
\label{sec:thermodynlim}
In the limit of an infinite lattice ($N\to \infty$) the free electronic band energies $\epsilon_{\mathbf{k}}$ become
continuous functions of the wavevector. All summations will therefore be replaced
by integrations:
\begin{equation}
\frac{1}{N}\sum_{\mathbf{k}}\rightarrow
\frac{1}{V_{\mathrm{BZ}}}\int\mathrm{d}\mathbf{k}.
\label{eq:sumtoint}
\end{equation}
In this limit all eigenstates with down-electron part whose energies lie in the
scattering region ($\epsilon_{\mathbf{k}}^{\mathrm{min}}-\frac{JS}{2}\le E \le
\epsilon_{\mathbf{k}}^{\mathrm{max}}-\frac{JS}{2}$) lose their norm
as becomes clear immediately by inspecting (\ref{eq:state_1_0}) and
(\ref{eq:norm_polstates}). Only the bound (polaron) state above ($J>0$) or below ($J<0$)
the scattering region will survive provided that $J$ is sufficiently large so
that the state is energetically well separated from the scattering states.\\
The question then becomes, what are the eigenstates in this limit?
We will construct scattering states and show, that these states (+ polaron state)
form a complete basis in the continuum limit. Thereafter we solve the
time-dependent Schr\"odinger equation which does not suffer from such difficulties
for arbitrary initial conditions and show the connection to Green's function
theory and the scattering states.
\subsection{scattering states}
In this section we ask for the result of a scattering process of an up-electron
with a magnon, that is we want to solve the Lippmann-Schwinger equation \cite{Lippmann1950}
:
\begin{equation}
|\mathbf{k-q}\uparrow;\mathbf{q}\rangle^{\pm} = |\mathbf{k-q}\uparrow;\mathbf{q}\rangle+
R_{\pm}H_1|\mathbf{k-q}\uparrow;\mathbf{q}\rangle^{\pm},
\label{eq:lippschwing}
\end{equation}
with the free propagator $R_{\pm}=(E_{0}-H_0\pm i0^+)^{-1}$.
First we divide the Hamiltonian (\ref{eq:hamiltonian}) into a free part $H_0$ and
an interaction part $H_1$:
\begin{eqnarray}
\label{eq:devideham}
H_0 &=& \sum_{\mathbf{k}\sigma}
\left(
	\epsilon_{\mathbf{k}}-z_{\sigma}\frac{JS}{2}
\right)\hat{n}_{\mathbf{k}\sigma} , \\
H_1 &=& -\frac{J}{2\sqrt{N}}\sum_{\mathbf{kq}\sigma}
\left( z_{\sigma}S_\mathbf{q}^{z}c_{\mathbf{k-q}-\sigma}^+c_{\mathbf{k}\sigma}
	+ S_\mathbf{q}^{\sigma}c_{\mathbf{k-q}-\sigma}^+c_{\mathbf{k}\sigma}
	\right) \nonumber \\
	&+&\sum_{\mathbf{k}\sigma}z_{\sigma}\frac{JS}{2}\hat{n}_{\mathbf{k}\sigma}. \nonumber
\end{eqnarray}
With this division one can derive:
\begin{eqnarray}
R_{\pm}H_1|\mathbf{k-q}\uparrow,\mathbf{q}\rangle &=& |Z_1\rangle+|Z_2\rangle \nonumber \\
R_{\pm}H_1|Z_1\rangle &=& A_1|Z_2\rangle, \nonumber \\
R_{\pm}H_1|Z_2\rangle &=& A_2(|Z_1\rangle+|Z_2\rangle),
\label{eq:iter_start}
\end{eqnarray}
where the $A_i$ and $|Z_i\rangle$ are given in the appendix (\ref{eq:def_Ai_Zi}).
Defining the matrix:
\begin{equation}
\hat{\mathbf{A}}=
\begin{pmatrix}
0   & A_1 \\
A_2 & A_2
\end{pmatrix}
=\hat{\mathbf{U}}\hat{\mathbf{D}}\hat{\mathbf{U}}^{-1},
\label{eq:A_Matrix}
\end{equation}
we can iterate (\ref{eq:lippschwing}) and get:
\begin{align}
|\mathbf{k}-\mathbf{q}\uparrow\rangle^{\pm} &=
	|\mathbf{k}-\mathbf{q}\uparrow \rangle + 
	\begin{pmatrix}1 \\ 1\end{pmatrix}
	\left(\sum_{l=0}^{\infty}\hat{\mathbf{A}}^l\right)
	\begin{pmatrix}|Z_1\rangle \\ |Z_2\rangle \end{pmatrix} \nonumber\\
&=	|\mathbf{k}-\mathbf{q}\uparrow \rangle + 
	\begin{pmatrix}1 \\ 1\end{pmatrix}
	\hat{\mathbf{U}}\left(\sum_{l=0}^{\infty}\hat{\mathbf{D}}^l\right)\hat{\mathbf{U}}^{-1}
	\begin{pmatrix}|Z_1\rangle \\ |Z_2\rangle \end{pmatrix} \nonumber \\
&=  |\mathbf{k}-\mathbf{q}\uparrow \rangle +
	\frac{|Z_1\rangle}{1-A_2(1+A_1)}+\frac{|Z_2\rangle}{\frac{1}{1+A_1}-A_2} \nonumber \\
&=	\alpha_{\mathbf{q}}^{\mathbf{k}}|\mathbf{k}\downarrow;0\rangle +
	\sum_{\mathbf{q'}}\beta_{\mathbf{q},\mathbf{q'}}^{\mathbf{k}}
	|\mathbf{k}-\mathbf{q'}\uparrow;\mathbf{q'}\rangle
\label{eq:scatt_states}	
\end{align}
for the ingoing and outgoing scattering states. The diagonal matrix $\hat{\mathbf{D}}$
and the matrix of the eigenvectors $\hat{\mathbf{U}}$ are given in the appendix
\ref{ap:scattsec} as equation (\ref{eq:mat_D_U}),
the coefficients $\alpha_{\mathbf{q}}^{\mathbf{k}}$ and $\beta_{\mathbf{q},\mathbf{q'}}^{\mathbf{k}}$
as equation (\ref{eq:coeff_scatt}). \\
It remains to show, that the so constructed scattering states form
a basis set in the limit of an infinite lattice. Applying the Hamiltonian results in:
\begin{eqnarray}
\label{eq:eigenval_scatt}
H|\mathbf{k}-\mathbf{q}\uparrow;\mathbf{q}\rangle^{\pm} & = &
\left(\epsilon_{\mathbf{k-q}}-\frac{JS}{2}\right)|\mathbf{k}-\mathbf{q}\uparrow;\mathbf{q}\rangle^{\pm}\\
&+& \sum_{\mathbf{q'}}\mathcal{R}_{\mathbf{q,q'}}^{\mathbf{k}}
|\mathbf{k}-\mathbf{q'}\uparrow;\mathbf{q'}\rangle.\nonumber
\end{eqnarray}
The residue term $\mathcal{R}_{\mathbf{q,q'}}^{\mathbf{k}}$ that is given in the
appendix (\ref{eq:restterm_eigenval}) goes to zero as $O(1/N)$.
Therefore the scattering states (\ref{eq:scatt_states}) indeed become the correct
eigenstates in the infinite lattice limit with densely lying eigenenergies in the expected
region. 
\subsection{time dependent Schr\"odinger equation}
Choosing the Ansatz:
\begin{equation}
|\Psi_{\mathbf{k}}(t)\rangle=\alpha_{\mathbf{k}}(t)|\mathbf{k}\downarrow;0\rangle
+\sum_{\mathbf{q}}\beta_{\mathbf{k,q}}(t)|\mathbf{k-q}\uparrow;\mathbf{q}\rangle
\label{eq:ansatz_wave_t}
\end{equation} 
for the wave-function one gets the following system of differential equations
from the time-dependent Schr\"odinger equation ($\hbar=1$):
\begin{eqnarray}
i\dot{\alpha}_{\mathbf{k}}(t)&=&\left( \epsilon_{\mathbf{k}}+\frac{JS}{2}\right)
\alpha_{\mathbf{k}}(t)
-J\sqrt{\frac{S}{2N}}\sum_{\mathbf{q}}\beta_{\mathbf{k,q}}(t),\nonumber \\
i\dot{\beta}_{\mathbf{k,q}}(t)&=&\left(\epsilon_{\mathbf{k-q}}-\frac{JS}{2}\right)
\beta_{\mathbf{k,q}}(t)-J\sqrt{\frac{S}{2N}}\alpha_{\mathbf{k}}(t) \nonumber \\
& & +\frac{J}{2N}\sum_{\mathbf{q}}\beta_{\mathbf{k,q}}(t). 
\label{eq:diff_equ_coeff}
\end{eqnarray}
This can be transformed into a system of algebraic equations by the Laplace transform:
\begin{equation}
\bar{f}(s)=\int_{0}^{\infty}\mathrm{e}^{-st}f(t)\mathrm{d}t;\:\mathrm{Re}(s)>0
\label{eq:def_laplacetr}
\end{equation}
and one obtains the solution for the coefficients in the $s$-domain:
\begin{eqnarray}
\label{eq:sol_s_space}
\bar{\alpha}_{\mathbf{k}}(s) &=& \frac{i\alpha_{\mathbf{k}}(0)}
	{is-\epsilon_{\mathbf{k}}-\frac{JS}{2}\left(1+
	\frac{JG^{(0)}(is+\frac{JS}{2})}{1-\frac{J}{2}G^{(0)}(is+\frac{JS}{2})}\right)} \\
&-& \frac{iJS}{\sqrt{2NS}}\sum_{\mathbf{q}}\frac{\beta_{\mathbf{k,q}}(0)}
	{\left( is-\epsilon_{\mathbf{k-q}}+\frac{JS}{2}\right)h_1(is)},\nonumber \\
\bar{\beta}_{\mathbf{k,q}}(s) &=&\frac{i\beta_{\mathbf{k,q}}(0)+\frac{i\alpha_{\mathbf{k}}(0)
	-(is-\epsilon_{\mathbf{k}}+\frac{JS}{2})\bar{\alpha}_{\mathbf{k}}(s)}{\sqrt{2NS}}}
	{is-\epsilon_{\mathbf{k-q}}+\frac{JS}{2}},
\end{eqnarray}
with
\begin{equation}
h_1(z) = z-\epsilon_{\mathbf{k}}-\frac{J}{2}
		 \left(S+1+\Phi_{\mathbf{k}}(z+\frac{JS}{2})\right).
\label{eq:h_1_z}
\end{equation}
For the back-transformation into the time-domain one has to solve the Bromwich integral:
\begin{equation}
f(t)=\frac{1}{2\pi i}\int_{\gamma-i\infty}^{\gamma+i\infty}\mathrm{e}^{st}\bar{f}(s)\mathrm{d}s,
\label{eq:bromwichint}
\end{equation}
where $\gamma$ is a real constant, that is larger than the real part of any singularity
of $\bar{f}(s)$. In our case all singularities/branch cuts of $\bar{\alpha}_{\mathbf{k}}(s)$
and $\bar{\beta}_{\mathbf{k,q}}(s)$ lie in a finite domain on the
imaginary axis. Therefore $\gamma$ can take any value $\gamma>0$.
\begin{figure}
    \includegraphics[width=0.6\linewidth]{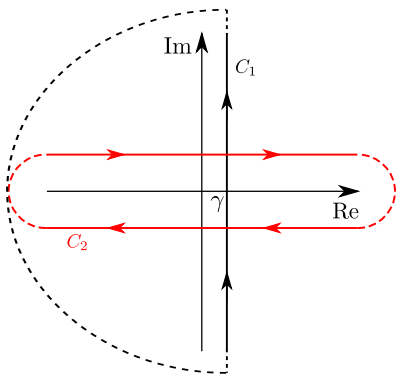}
    \caption{\label{fig3}(color online) Integration contour in complex plane.}
\end{figure}
The integration path is shown in Fig.~\ref{fig3} and denoted by $C_1$. The contour
can be closed at infinity since the integrand vanishes there (dashed black line).
Performing the variable substitution $s\to-iz$ and a contour deformation one can
change the integration path to become $C_2$ (red line). The singularities/
branch cuts now lie on the real axis.\\
We will give the solutions for two special boundary conditions here. The choice:
\begin{equation}
\mathrm{a)}\;\alpha_{\mathbf{k}}(t=0)=1; \beta_{\mathbf{k,q}}(t=0)=0,\forall \mathbf{q}
\label{eq:boundcond_a}
\end{equation}
results in:
\begin{align}
\label{eq:a_k_bounda}
\alpha_{\mathbf{k}}(t) &= \frac{-1}{2\pi i}\int_{C_2}\mathrm{d}z
\frac{\mathrm{e}^{-izt}}{z-\epsilon_{\mathbf{k}}-\frac{JS}{2}
    \left(
        1+\frac{JG^{(0)}(z+\frac{JS}{2})}{1-\frac{J}{2}G^{(0)}(z+\frac{JS}{2})}
    \right)} \nonumber \\
    &= \frac{-1}{2\pi i}\int_{-\infty}^{\infty}\mathrm{d}x\: \mathrm{e}^{-ixt}
    \left( G_{\mathbf{k}\downarrow}(x+i0^+) - G_{\mathbf{k}\downarrow}(x-i0^+)
    \right) \nonumber \\
    &= \int_{-\infty}^{\infty}\mathrm{d}x\: \mathrm{e}^{-ixt}
        S_{\mathbf{k}\downarrow}(x),
\end{align}
that is, $\alpha_{\mathbf{k}}(t)$ is the Fourier transform of the spectral density
obtained from the down-electron Green's function derived earlier (\ref{eq:Green_down}).\\
Choosing: 
\begin{equation}
\mathrm{b)}\; \beta_{\mathbf{k,q_n}}(0)=1;\alpha_{\mathbf{k}}(0)=
\beta_{\mathbf{k,q_m}}(0)=0,\forall m\ne n
\label{eq:boundcond_b}
\end{equation}
one gets:
\begin{equation}
\alpha_{\mathbf{k}}(t) = \frac{JS}{\sqrt{2NS}}\frac{1}{2\pi i}\int_{C_2}\mathrm{d}z\:
\frac{\mathrm{e}^{-izt}}{\left( z -\epsilon_{\mathbf{k-q_n}}+\frac{JS}{2}\right)
		h_1(z)}.
\label{eq:alpha_bound_b}
\end{equation}
It is to be expected, that this will become the $\alpha_{\mathbf{q}_n}^{\mathbf{k}}$ of
the scattering states (\ref{eq:scatt_states}) for large times. We will test this
conjecture in the next section, when we have build up the machinery to investigate
different time domains of $\alpha_{\mathbf{k}}(t)$. 
\section{dynamics of decay}
\label{sec:dod}
\begin{figure}
    \includegraphics[width=0.95\linewidth]{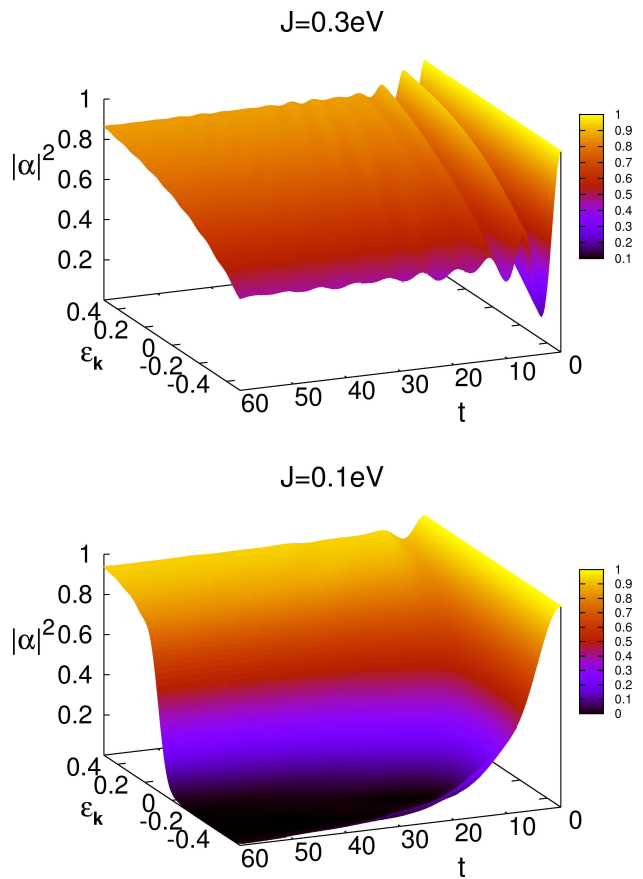}
    \caption{\label{fig4}(color online) Probability to find a down-electron
    $P(t)=|\alpha_{\mathbf{k}}(t)|^2$ as a function of time and the free electron
    dispersion $\epsilon_{\mathbf{k}}$ for two different values of $J$. The time
    is given in units of $\approx 6.58212*10^{-16}$s.
    Parameters: $S=3.5$, $W=1.0$ eV.}
\end{figure}

The absolute square of the coefficient $\alpha_{\mathbf{k}}(t)$ in
(\ref{eq:ansatz_wave_t}) can be interpreted as the probability to find a
down-electron in the system at time $t$. In Fig.~\ref{fig4} we show
$P(t)=|\alpha_{\mathbf{k}}(t)|^2$ of a system subjected to the boundary conditions
(\ref{eq:boundcond_a}) for two different couplings $J=0.1/0.3$ eV.
For large $J$ (upper panel in Fig.~\ref{fig1} shows the corresponding spectral
density) there is always a finite probability to find a down-electron in
the system. $P(t)$ shows characteristic oscillations over time. These oscillation
will be damped with increasing $t$ and $P(t)$ becomes static in the long-run limit.
When $J$ is smaller, parts of the polaron band dip into the scattering states
(lower panel in Fig.~\ref{fig1}). For wavevectors $\mathbf{k}$ where 
$\epsilon_{\mathbf{k}}$ lies below a certain threshold $P(t)$ decreases
exponentially and become zero in the long-run limit. The down-electron decays
inevitably. However, it is clearly visible in the figure, that there are
deviations from an exponential decay law in the short and the long time regime. 
We will now give a more detailed analysis of the different time domains.
\subsection{short time behavior}
For small times a Taylor expansion of the exponential function in
(\ref{eq:a_k_bounda}) gives:
\begin{eqnarray}
\alpha_{\mathbf{k}}(t) &=&
\int\limits_{-\infty}^{\infty}\mathrm{d}xS_{\mathbf{k}\downarrow}(x)\left[1+ixt-\frac{(xt)^2}{2}
-i\frac{(xt)^3}{6}+\cdots\right]\nonumber\\
&=& 1+iM_{\mathbf{k}\downarrow}^{(1)}t
-\frac{M_{\mathbf{k}\downarrow}^{(2)}}{2}t^2-i\frac{M_{\mathbf{k}\downarrow}^{(3)}}{6}t^3
+O(t^4).
\label{eq:a_k_short_t}
\end{eqnarray}
The spectral moments $M_{\mathbf{k}\downarrow}^{(i)}$ can be calculated exactly
in our case and we give the first four in the appendix (\ref{app:spectral_m_down}).
From this we get the short time expansion of $P(t)$:
\begin{eqnarray}
|\alpha_{\mathbf{k}}(t)|^2 = 1-C_1t^2 + O(t^4)
\label{eq:P_short_t}
\end{eqnarray}
with
\begin{equation}
C_1=M_{\mathbf{k}\downarrow}^{(2)}-(M_{\mathbf{k}\downarrow}^{(1)})^2 =
\frac{J^2S}{2}.
\label{eq:c1_P_short_t}
\end{equation}
Since $C_1\ge0$ $P(t)$ will always decrease (or stay constant) for small times
as it should be for normalization reasons. The change of $P(t)$ has a zero slope
at $t=0$ since $\mathrm{d}P(t)/\mathrm{d}t|_0=0$.
This leads to the famous quantum Zeno effect\cite{Misra1977} - a down electron
whose existence is monitored continuously by measurement\footnote{``Continuously'' 
means in sufficient short time intervals so that $C_1t^2 \ll 1$.} will
never decay. 
\subsection{intermediate and long time behavior}
\begin{figure}
    \includegraphics[width=0.95\linewidth]{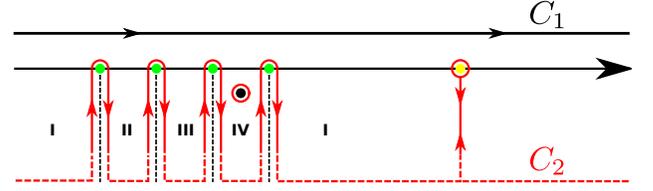}
    \caption{\label{fig5}(color online) Deformation of the integration contour in the complex
    plane to study the intermediate and long time behavior. Green dots: van Hove
    singularities of the scattering states. Yellow dot: separated polaron pole.
    Black dot: complex pole of decaying state.}
\end{figure}

\begin{figure}
    \includegraphics[width=0.95\linewidth]{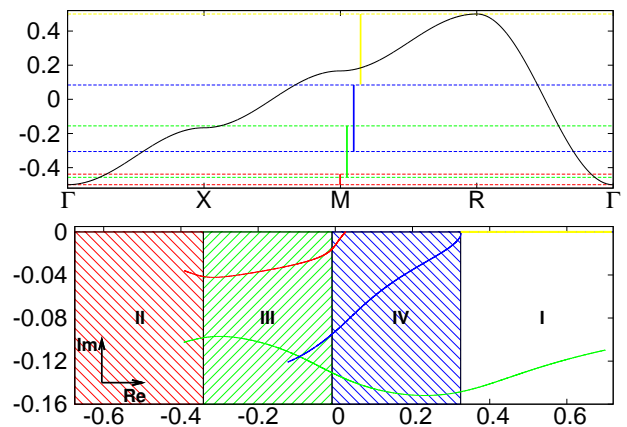}
    \caption{\label{fig6}(color online) Lower figure: Pole lines (parametrized
    by $\epsilon_{\mathbf{k}}$) in the different Riemann
    sheets. The colors of the lines denote the sheet where the poles can be
    found: red: II, green: III, blue: IV. When a pole lies in its window of
    visibility (dashed area with the same color) it gives an exponentially
    declining contribution to $\alpha_{\mathbf{k}}(t)$. The yellow line shows
    the positions of the polaron peak, when it lies outside the scattering
    region. Upper figure: Free electron dispersion $\epsilon_{\mathbf{k}}$
    and the windows of visibility for the different poles.
    Parameters: $S=3.5$, $J=0.1$ eV and $W=1.0$ eV.}
\end{figure}

\begin{figure}
    \includegraphics[width=0.95\linewidth]{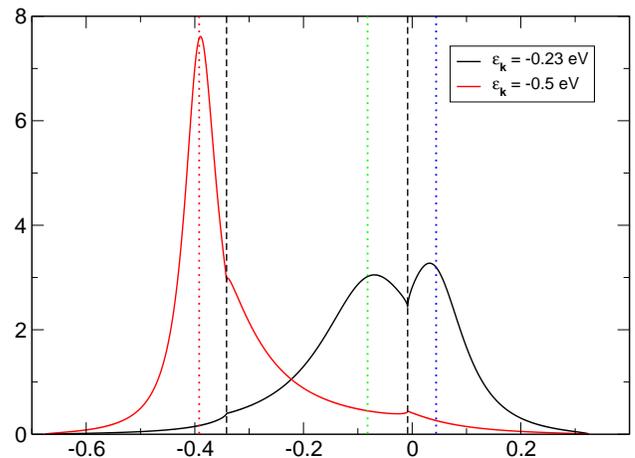}
    \caption{\label{fig9}(color online) Down-electron spectral density for two different values
    of $\epsilon_{\mathbf{k}}$. Vertical black, dashed lines: position of the
    van Hove singularities. Vertical dotted lines: real part of the pole
    positions. Parameters: $S=3.5$, $J=0.1$ eV and $W=1.0$ eV.}
\end{figure}

For times $t>0$ we can omit the part of the integration contour $C_2$ in (\ref{eq:a_k_bounda}) 
that lies below the real axis because it does not contribute to the integral. 
The remaining integration path is shown in Fig.~\ref{fig5} as $C_1$. To study
the long time behavior we pull the contour into the lower complex plane.
For this aim we mention first, that the down electron GF has a branch cut in the
energy region of the scattering states. To pull the contour below the real axis
we have to find the analytic continuation of the GF over the cut from above to
below the axis. This is equivalent to find the analytic continuation of the free
GF: $G^{(0)}(z)$ as becomes clear by inspecting (\ref{eq:a_k_bounda}).
To be specific we choose here the free GF of the simple cubic lattice with
the dispersion given in (\ref{eq:dispersion}). The analytic continuation of this
function is discussed in detail in the appendix (\ref{ap:analytcont}). The
important point is, that the van Hove singularities are branch points and one 
reaches different Riemann sheets depending on the
position relative to the singularity (left/right) where the continuation is
performed. \\ The situation is depicted in Fig.~\ref{fig5}.
The green dots mark the positions of the van Hove singularities. The deformed
contour $C_2$ (red) cannot overcome these points analytically. The contributions
to the integral (\ref{eq:a_k_bounda}) from the path left and right of the
singularities will not cancel each other because they lie on different sheets of
the Riemann surface. We exemplify the (approximate) analysis of the integral by giving
a concrete expression for the path around the lower band edge at
$x_{\mathrm{s}}^{\mathrm{I/II}}=-\left(\frac{JS}{2}+\frac{W}{2}\right)$. 
\begin{eqnarray}
\lefteqn{\alpha_{\mathbf{k}}^{\mathrm{I/II}}(t) = 
\frac{-1}{2\pi}\int\limits_{-\infty}^{0}\mathrm{d}x\:\mathrm{e}^{-i\left(x_{\mathrm{s}}^{\mathrm{I/II}}+ix\right)t}} \\ 
&&\times
\left(
    G_{\mathbf{k}\downarrow}^{\mathrm{I}}\left(x_{\mathrm{s}}^{\mathrm{I/II}}+ix\right) 
   -G_{\mathbf{k}\downarrow}^{\mathrm{II}}\left(x_{\mathrm{s}}^{\mathrm{I/II}}+ix\right) 
\right) \nonumber \\
&=& 
\frac{-1}{2\pi}\int\limits_{-\infty}^{0}\mathrm{d}x\:\mathrm{e}^{-i\left(x_{\mathrm{s}}^{\mathrm{I/II}}+ix\right)t}
\left(C_s^{\mathrm{I/II}}\sqrt{x}+\mathcal{O}\left(
x^{\frac{3}{2}}\right)\right) \nonumber \\
&=&\frac{-i}{2\sqrt{\pi}}\mathrm{e}^{-ix_{\mathrm{s}}^{\mathrm{I/II}}t}\left(\frac{1}{2}C_s^{\mathrm{I/II}}t^{-\frac{3}{2}}
    +\mathcal{O}(t^{-\frac{5}{2}})\right).
\label{eq:alpha_I_II}
\end{eqnarray}
The other van Hove singularities give rise to similar terms and we have summarized
the explicit expressions for the coefficients $C_s$ in the appendix (\ref{ap:coeff_long_t}).
From (\ref{eq:alpha_I_II}) we see, that the van Hove singularities contribute an oscillatory
term with a frequency equal to the energetic position to $\alpha_{\mathbf{k}}(t)$. This
term is damped in time with a decay rate given by a power law. The exponent of this power law
depends only on the nature of the van Hove singularity, that is a square root singularity in
our case\cite{Hoehler1958}.\\
For large enough $J$ there is a separated pole of the down-electron
GF on the real axis shown as yellow dot in Fig.~\ref{fig5}. If we denote the position of this
pole by $x_{\mathrm{pol}}$ the contribution to $\alpha_{\mathbf{k}}(t)$ is given as the residue 
of the integrand in (\ref{eq:a_k_bounda}) at this point:
\begin{eqnarray}
\alpha_{\mathbf{k}}^{\mathrm{pol}}(t)&=&\left.\frac{\mathrm{e}^{-ixt}}
	{1-JS\partial_x\left(\frac{1}
	{1-\frac{J}{2}G^{(0)}\left(x+\frac{JS}{2}\right)}\right)}\right|_{x=x_{\mathrm{pol}}}
	\nonumber \\
	&=& C_{\mathrm{pol}}\mathrm{e}^{-ix_{\mathrm{pol}}t}.
\label{eq:alpha_pol}
\end{eqnarray}
The characteristic oscillations of $P(t)$ visible in the upper panel of Fig.~\ref{fig4}
are superpositions of the oscillatory terms in (\ref{eq:alpha_I_II}) and (\ref{eq:alpha_pol}).
They are damped in time as ($\sim t^{-\frac{3}{2}}$) at least. For large times
only the contribution from (\ref{eq:alpha_pol}) will survive and we get:
\begin{equation}
\lim_{t\to\infty}|\alpha_{\mathbf{k}}(t)|^2=|C_{\mathrm{pol}}|^2.
\label{eq:alphasq_larget}
\end{equation}
For large coupling strength $|J|\gg1$ the pole position becomes approximately
$x_{\mathrm{pol}}\approx\frac{J}{2}(S+1)$. Using the high energy expansion
of the free GF in (\ref{eq:alpha_pol}) we can derive an explicit result for
$P(t)$ in this limit:
\begin{equation}
\lim_{t\to\infty}|\alpha_{\mathbf{k}}(t)|^2\stackrel{|J|\gg1}{\approx}\left(\frac{2S}{2S+1}\right)^2.
\label{eq:alphasq_lt_lJ}
\end{equation}
The electron polarization can be calculated from $P(t)$:
\begin{equation}
\corr{\sigma_z}(t)=\frac{\sum_{\mathbf{q}}|\beta_{\mathbf{k,q}}(t)|^2
-|\alpha_{\mathbf{k}}(t)|^2}{2}=\frac{1}{2}-|\alpha_{\mathbf{k}}(t)|^2,
\label{eq:electron_polariz}
\end{equation}
where in the last step the normalization condition $\sum_{\mathbf{q}}|\beta_{\mathbf{k,q}}(t)|^2
+|\alpha_{\mathbf{k}}(t)|^2=1$ is used. \\
We come now to the case of small $|J|$, when parts of the polaron band lie
inside the scattering states. For those $\mathbf{k}$-values the initial down-electron
will decay over time and vanish completely in the long time limit.
One therefore expects an exponential decaying contribution to
$\alpha_{\mathbf{k}}(t)$. As soon as the polaron peak runs into the scattering
states there appear poles in regions II,III or IV of the analytic continuation of
the down-electron GF that account for such an exponential term. This is depicted 
in Fig.~\ref{fig5} by the black dot in region IV. These poles contribute a term
to $\alpha_{\mathbf{k}}(t)$ that equals (\ref{eq:alpha_pol}) but now with a
complex pole energy $x_{\mathrm{pol}}=x_r-i\gamma$.\\
There is one pole in each sheet of the different analytic continuations. In
Fig.~\ref{fig6} (lower panel) the pole lines in the complex plane (parametrized by
$\epsilon_{\mathbf{k}}$) are shown for $J=0.1$ eV. Whenever a pole has a real
part $x_r$, that lies inside the energy region where the analytic continuation
of the corresponding sheet has been performed (II, III or IV) it will add a term
(\ref{eq:alpha_pol}) to $\alpha_{\mathbf{k}}(t)$. The upper panel of
Fig.~\ref{fig6} shows the free dispersion $\epsilon_{\mathbf{k}}$ for a path
along the high symmetry lines in the first Brillouin zone. 
The $\epsilon_{\mathbf{k}}$ for which the corresponding pole is inside its region
of visibility can be read off from the horizontal colored lines. Whenever this is
the case, a pronounced (quasi particle) peak is visible in the down-electron spectral density,
that lies approximately at the position of the real part of the pole $x_r$. We
demonstrate this in Fig.~\ref{fig9} for two different values of
$\epsilon_{\mathbf{k}}$. 
For $\epsilon_{\mathbf{k}}=-0.5$ eV there is only a pole in region II (endpoint of
the red pole-line). One finds a distinctive quasi-particle peak in the spectral
density (red line) at approximately the real part of the pole position
(vertical dotted line). An interesting constellation arises for
$\epsilon_{\mathbf{k}}=-0.23$ eV. From Fig.~\ref{fig6} one can read off that two
poles are now in their respective region (III,IV). Consequently two
quasi-particle peaks can be found in the SD (black line).
The negative imaginary part of the pole position can be
interpreted as (half of) the inverse lifetime: $\tau\sim\frac{2}{\gamma}$.
This is reflected directly by the width of the quasi-particle peaks, the smaller
$\tau$ the broader the corresponding peak in the SD.\\
We had seen, that there are deviations from an exponential decay law at small
times. The same is true for the long time limit. When $t\gg\frac{2}{\gamma}$ the
exponentially decaying term (\ref{eq:alpha_pol}) is damped away and only the
contribution from the van Hove singularities (\ref{eq:alpha_I_II}) remain, which
leads to a power law decay rate ($|\alpha_{\mathbf{k}}(t)|^2\sim t^{-3}$). 
It was first shown by Khalfin\cite{Khalfin1958}, that this deviation from
exponential decay at large times is a general property of a system with a
spectrum, that is bounded from below.\\
At the end of this section we want to discuss the long time behavior of a
system, where we have chosen the initial conditions (\ref{eq:boundcond_b}). 
If we deform the integration contour $C_2$ in (\ref{eq:alpha_bound_b}) as
shown in Fig.~\ref{fig5} and chose $J$ sufficiently small, so that $h_1(z)$
has no real root outside the scattering spectrum we get for large times:
\begin{eqnarray}
\label{eq:bound_b_alpha_long_t}
\lefteqn{\lim_{t\to\infty}\alpha_{\mathbf{k}}(t)=\mathrm{e}^{-i(\epsilon_{\mathbf{k-q_n}}-\frac{JS}{2})t}\times}\\
&&\frac{1}{\sqrt{2NS}}
\frac{-JS}{\epsilon_{\mathbf{k-q}}-\epsilon_{\mathbf{k}}-\frac{J}{2}(2S+1)
	-\frac{J}{2}\Phi_{\mathbf{k}}(\epsilon_{\mathbf{k-q}}+i0^+)}.\nonumber
\end{eqnarray}
This is indeed identical to the $\alpha_{\mathbf{q}}^{\mathbf{k}}$ coefficient found earlier
for the scattering states (\ref{eq:scatt_states}) up to a trivial time dependent factor.
For larger $J$ there will be additional terms coming from the real root of
$h_1(z)$ which are not contained in the result of scattering theory. The reason
for this are the persistent self-energy effects (cloud effects) already
mentioned by van Hove\cite{Hove1955}.
\section{Summary and Conclusions}
\label{sec:sumcon}
In this work we have revised the theory of the magnetic polaron to get a deeper
understanding of one of the few exactly solvable limiting cases of the KLM.

First we have derived the complete eigenvalue spectrum of the finite lattice and
the corresponding eigenstates are constructed. 

In the limit of an infinite lattice parts of these eigenstates lose their norm and
the eigenvalues of these states degenerate with the eigenvalues of the pure
up-states.
We then calculate the scattering states and show, that these states become the new
eigenstates in this limit. 

By solving the time-dependent Schr\"odinger equation for arbitrary initial
conditions we are able to give a detailed analysis of the decay dynamics of the
system. This is done for a down-electron with wavevector $\mathbf{k}$ prepared
at $t=0$. For large exchange coupling $J$ the probability $P(t)$ to find a
down-electron at time $t>0$ shows characteristic oscillations which are damped as
$\sim t^{-\frac{3}{2}}$. We find the reason of these oscillations to be
interference of different oscillatory terms with frequencies
equal to the energetic positions of the van Hove singularities and the separated
pole (polaron peak). For large times $P(t)$ becomes static with a value $P(t)<1$
and for $|J|\gg1$ an explicit value can be derived.

For small coupling $J$ and $\epsilon_{\mathbf{k}}$ below a certain threshold
$P(t)$ goes to zero over time - the down-electron decays. The main contribution
to the decay dynamics stems from poles in the analytic continuation of the
propagator giving rise to an exponential decay law. The imaginary part of the
poles determines the lifetime of the down-electron.

We find deviations from an exponential decay law for small and large times.
At $t=0$ the slope of $P(t)$ is zero. This leads to the quantum Zeno effect.
At large times the exponential term is damped away and the decay behavior is
solely determined by the contributions from the van Hove singularities. We get
a power law decay ($P(t)\sim t^{-3}$) in this regime and there are also
oscillations caused by interference effects.

The connection between certain initial conditions for the solution of the time-dependent
Schr\"odinger equation and the retarded Green's function approach as well as
scattering theory is worked out.
When a pure down-electron state is prepared at $t=0$ the time development of the
down state coefficient of the wave function is given by the time-dependent
down-electron spectral density, that can be obtained from the retarded Green's
function.
Preparing an up-electron and a magnon at $t=0$ and taking the $t\to\infty$ limit
the result of scattering theory (for the down-electron coefficient) can be
reproduced for small $J$. For large $J$ however a new quasi-particle with
infinite lifetime is created. Scattering theory is not able to describe this
formation of a new quasi-particle.

The limiting case of the magnetic polaron is believed to describe the essential
physics of the magnetic semiconductors EuO and EuS. For a realistic description
of the quasi particle density of states of these materials one has to replace the model band
structure (\ref{eq:dispersion}) used in this work by a material specific one,
that can be obtained from a ab initio band structure calculation. Such a
calculation was done by Nolting et. al.\cite{Nolting1987} for EuO and Borstel
et. al.\cite{Borstel1987} for EuS. From the experimental side the natural way of
measuring the unoccupied states would be an inverse photoemission experiment.
An alternative method is the two-photon photoemission spectroscopy\cite{Steinmann1989}.
This experimental technique can be used to get time and spin resolved spectral
information with time resolutions down to the femtosecond and even attosecond
regime \cite{Scholl1997, Pickel2006,Cavalieri2007}. These new experimental
methods could open up a possible road to a direct measurement of the decay
dynamics of an (down) electron in the above mentioned prototypical compounds.

\appendix
\section{Definitions scattering section}
\label{ap:scattsec}
\begin{eqnarray}
A_1 &=& \frac{JS}{\epsilon_{\mathbf{k-q}}-\epsilon_{\mathbf{k}}-JS\pm i0^+}, \nonumber \\
A_2 &=& \frac{J}{2N}\sum_{\mathbf{q'}}\frac{1}
		 {\epsilon_{\mathbf{k-q}}-\epsilon_{\mathbf{k-q'}}\pm i0^+}, \nonumber \\
|Z_1\rangle & = & -\frac{J}{2}\sqrt{\frac{2S}{N}}\frac{|\mathbf{k}\downarrow;0\rangle}
	{\epsilon_{\mathbf{k-q}}-\epsilon_{\mathbf{k}}-JS\pm i0^+}, \nonumber \\
|Z_2\rangle & = & \frac{J}{2N}\sum_{\mathbf{q'}}	
	\frac{|\mathbf{k-q'}\uparrow;\mathbf{q'}\rangle}
		 {\epsilon_{\mathbf{k-q}}-\epsilon_{\mathbf{k-q'}}\pm i0^+}.
\label{eq:def_Ai_Zi}
\end{eqnarray}
\begin{align}
\hat{\mathbf{D}} &= \frac{A_2}{2}
\begin{pmatrix}
1-\sqrt{4\frac{A_1}{A_2}+1}& 0 \\
0& 1+\sqrt{4\frac{A_1}{A_2}+1}
\end{pmatrix}, \\
\hat{\mathbf{U}} &=
\begin{pmatrix}
-\frac{1}{2}-\sqrt{\frac{A_1}{A_2}+\frac{1}{4}} & -\frac{1}{2}+\sqrt{\frac{A_1}{A_2}+\frac{1}{4}} \\
1 & 1
\end{pmatrix}.
\label{eq:mat_D_U}
\end{align}

\begin{align}
\alpha_{\mathbf{q}}^{\mathbf{k}} &=
	\frac{1}{\sqrt{2NS}}\frac{-JS}{\epsilon_{\mathbf{k-q}}-\epsilon_{\mathbf{k}}-\frac{J}{2}(2S+1)
	-\frac{J}{2}\Phi_{\mathbf{k}}(\epsilon_{\mathbf{k-q}} \pm i0^+)}, \nonumber \\
\beta_{\mathbf{q},\mathbf{q'}}^{\mathbf{k}} &= \delta_{\mathbf{q,q'}}
	-\frac{1}{\sqrt{2NS}}
	\left( 1+\frac{\epsilon_{\mathbf{k-q'}}-\epsilon_{\mathbf{k}}}
		          {\epsilon_{\mathbf{k-q}}-\epsilon_{\mathbf{k-q'}}\pm i0^+}
	\right) \alpha_{\mathbf{q}}^{\mathbf{k}}.
\label{eq:coeff_scatt}
\end{align}

\begin{eqnarray}
\mathcal{R}_{\mathbf{q,q'}}^{\mathbf{k}} &=& \frac{\alpha_{\mathbf{q}}^{\mathbf{k}}}
{\sqrt{2NS}}
\left\{
	\frac{2(\epsilon_{\mathbf{k-q}}-\epsilon_{\mathbf{k-q'}})
		(\epsilon_{\mathbf{k-q}}-\epsilon_{\mathbf{k}})}
		{\epsilon_{\mathbf{k-q}}-\epsilon_{\mathbf{k-q'}}\pm i0^+}\right. \nonumber \\
&& \left.\pm i0^+ \frac{2\epsilon_{\mathbf{k-q}}-\epsilon_{\mathbf{k-q'}}-\epsilon_{\mathbf{k}}}
		{\epsilon_{\mathbf{k-q}}-\epsilon_{\mathbf{k-q'}}\pm i0^+}
\right\}
\label{eq:restterm_eigenval}
\end{eqnarray}

\section{Analytic continuation of $G^{(0)}(z)$}
\label{ap:analytcont}
The simple cubic lattice Green's function can be expressed in terms of the
Gauss hypergeometric function\cite{Joyce2004}:
\begin{equation}
G^{(0)}(z)=\mu_1(z)\left({}_2F_1\left(\frac{1}{3},\frac{2}{3};1;\eta_1(z)\right)\right)^2
\label{eq:green_0}
\end{equation}
with
\begin{eqnarray}
\label{eq:mu1_eta1}
\mu_1(z) & = &
\frac{1}{2z}\left(3\sqrt{1-\frac{1}{z^2}}-\sqrt{1-\frac{9}{z^2}}\right), \\
\eta_1(z) & = &
\frac{1}{8z^2}\left(4z^2+(9-4z^2)\sqrt{1-\frac{9}{z^2}}-27\sqrt{1-\frac{1}{z^2}}\right). \nonumber
\end{eqnarray}
This function is analytic in the complete complex plane with the exception of a
branch cut on the real axis in the region $E_B\in-3\dots3$ (bandwidth: $W=6$). 
When the real axis is crossed in this region, the imaginary part of
(\ref{eq:green_0}) shows a discontinuity of $\Delta(x)=\pm2\pi \rho(x)$ where
$\rho(x)$ denotes the free density of states at point $x$. 
\begin{figure}
    \includegraphics[width=0.9\linewidth]{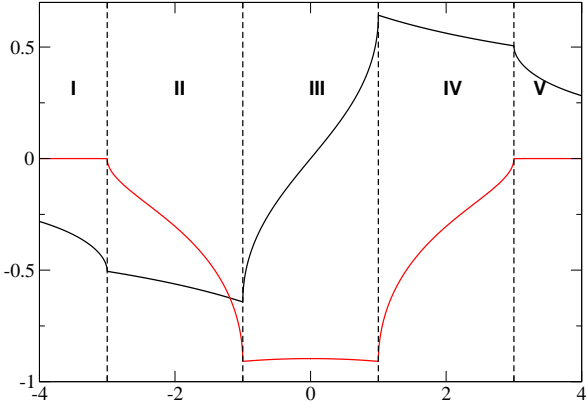}
    \caption{\label{fig7}(color online) Free GF slightly above the real axis. Vertical dashed
    lines: position of the van Hove singularities that form the boundaries for
    a possible analytic continuation below the real axis. Parameters: $W=1.0$
    eV.}
\end{figure}
In Fig.~\ref{fig7} $G^{(0)}(x+i0^+)$ is plotted slightly above the real axis.
The clearly visible van Hove singularities at the band edges ($-3,3$) and at 
($-1,1$) are limiting points for a possible analytic continuation over the branch
cut. Therefore one reaches different Riemann sheets, depending on the section
(II, III or IV) where the analytic continuation is performed. \\
To find these continuations we discuss firstly the analytic properties of the
constituting functions. The square root function has branch points (of first
order) at $0,-\infty$
and a branch cut connecting them on the negative real axis. Whenever one crosses
this cut, one has to change the sign of the square root function to get an
analytic continuation. \\ The hypergeometric function:
\begin{equation}
\mathrm{Hyp1}(z) = {}_2F_1\left( \frac{1}{3},\frac{2}{3};1;x\right)
\label{eq:hyp1}
\end{equation}
has a branch point of second order at the points $1,\infty$ and a branch cut
on the real axis connecting them. Using formulas given in
[\onlinecite{Abramowitz1964}] one finds for the continuation from above to
below the real axis:
\begin{eqnarray}
\label{eq:hyp2}
\mathrm{Hyp2}(z) & = & (-z)^{-\frac{1}{3}}\frac{\Gamma\left(
    \frac{1}{3}\right)\mathrm{e}^{-i\frac{4\pi}{3}}}
    {\left(\Gamma\left( \frac{2}{3}\right)\right)^2}
    \:{}_2F_1\left(\frac{1}{3},\frac{1}{3};\frac{2}{3};\frac{1}{z}\right)
    \\
    & + & (-z)^{-\frac{2}{3}}\frac{\Gamma\left(
    -\frac{1}{3}\right)\mathrm{e}^{-i\frac{8\pi}{3}}}
    {\left(\Gamma\left( \frac{1}{3}\right)\right)^2}
    \:{}_2F_1\left(\frac{2}{3},\frac{2}{3};\frac{4}{3};\frac{1}{z}\right),\nonumber
\end{eqnarray}
and from below to above the real axis:
\begin{eqnarray}
\label{eq:hyp3}
\mathrm{Hyp3}(z) & = & (-z)^{-\frac{1}{3}}\frac{\Gamma\left(
    \frac{1}{3}\right)\mathrm{e}^{-i\frac{2\pi}{3}}}
    {\left(\Gamma\left( \frac{2}{3}\right)\right)^2}
    \:{}_2F_1\left(\frac{1}{3},\frac{1}{3};\frac{2}{3};\frac{1}{z}\right)
    \\
    & + & (-z)^{-\frac{2}{3}}\frac{\Gamma\left(
    -\frac{1}{3}\right)\mathrm{e}^{-i\frac{4\pi}{3}}}
    {\left(\Gamma\left( \frac{1}{3}\right)\right)^2}
    \:{}_2F_1\left(\frac{2}{3},\frac{2}{3};\frac{4}{3};\frac{1}{z}\right).\nonumber
\end{eqnarray}
With this analysis it is now easy to find the analytic continuation of
(\ref{eq:green_0}) from above to below the branch cut. To exemplify this, we
discuss the continuation in region II. By crossing the real axis one also crosses
the branch cut of the square root function $\sqrt{1+\frac{3}{x}}$. Defining new
functions:
\begin{eqnarray}
\label{eq:mu2_eta2}
\mu_2(z) & = &
\frac{1}{2z}\left(3\sqrt{1-\frac{1}{z^2}}+\sqrt{1-\frac{9}{z^2}}\right), \\
\eta_2(z) & = &
\frac{1}{8z^2}\left(4z^2-(9-4z^2)\sqrt{1-\frac{9}{z^2}}-27\sqrt{1-\frac{1}{z^2}}\right), \nonumber
\end{eqnarray}
one then finds:
\begin{equation}
G_{\mathrm{II}}^{(0)}(z)=\mu_2(z)\left(\mathrm{Hyp1}(\eta_2(z)) \right)^2
\label{eq:analyt_cont_II}
\end{equation}
for the analytic continuation.\\
\begin{figure}
    \includegraphics[width=0.9\linewidth]{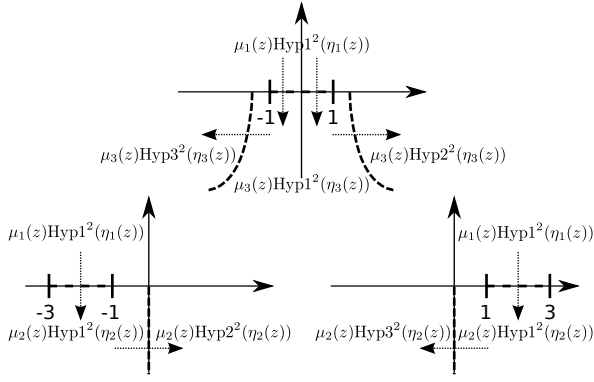}
    \caption{\label{fig8}Analytic continuations of the free GF from above to
    below the real axis over the different branch cuts. Parameters: $W=1.0$ eV.}
\end{figure}
Similar results can be obtained for the other regions. They are summarized in
Fig.~\ref{fig8}. The two missing definitions are:
\begin{eqnarray}
\label{eq:mu3_eta3}
\mu_3(z) & = &
-\frac{1}{2z}\left(3\sqrt{1-\frac{1}{z^2}}-\sqrt{1-\frac{9}{z^2}}\right), \\
\eta_3(z) & = &
\frac{1}{8z^2}\left(4z^2-(9-4z^2)\sqrt{1-\frac{9}{z^2}}+27\sqrt{1-\frac{1}{z^2}}\right). \nonumber
\end{eqnarray}
\section{spectral moments}
\label{app:spectral_m_down}
The moments of the down-electron spectral density are defined by:
\begin{equation}
M_{\mathbf{k}\downarrow}^{(n)} = \int\limits_{-\infty}^{\infty}\mathrm{d}E E^n 
S_{\mathbf{k}\downarrow}(E).
\label{eq:spect_moment_down}
\end{equation}
They can be calculated algebraically by the following rule\cite{Nolting2009}:
\begin{equation}
M_{\mathbf{k}\downarrow}^{(n)} = \corr{[\underbrace{[\dots[[c_{\mathbf{k}\downarrow},H]_{-},H]_{-}\dots,H]_{-}}_{n-\mathrm{times}},c_{\mathbf{k}\downarrow}^{+}]_+}.
\label{eq:spect_moment_down_calc}
\end{equation}
We give the first four moments for the limiting case of the magnetic polaron:
\begin{eqnarray}
M_{\mathbf{k}\downarrow}^{(0)} &=& 1, \\
\label{eq:zero_d_m}
M_{\mathbf{k}\downarrow}^{(1)} &=& \epsilon_{\mathbf{k}}+\frac{JS}{2}, \\
\label{eq:first_d_m}
M_{\mathbf{k}\downarrow}^{(2)} &=& \left(\epsilon_{\mathbf{k}}+\frac{JS}{2}\right)^2+\frac{J^2S}{2},\\
M_{\mathbf{k}\downarrow}^{(3)} &=& \left(\epsilon_{\mathbf{k}}+\frac{JS}{2}\right)^3\nonumber\\
&&    + J^2S(\epsilon_{\mathbf{k}}+J(S+1)/4). 
\end{eqnarray}
\begin{widetext}
\section{coefficients long time expansion}
\label{ap:coeff_long_t}
\begin{eqnarray}
x_s^{\mathrm{I/II}}&=&-\left( \frac{JS}{2}+\frac{1}{2}\right):\nonumber \\
C_s^{\mathrm{I/II}}&=&-\frac{4 (-1)^{3/4} \text{Hyp1}\left(\frac{1}{4} \left(2-\sqrt{2}\right)\right) J^2 S
   \left(3 \sqrt{2} \text{Hyp1}'\left(\frac{1}{4} \left(2-\sqrt{2}\right)\right)+2
   \text{Hyp1}\left(\frac{1}{4} \left(2-\sqrt{2}\right)\right)\right)}{\left((2
   \epsilon_{\mathbf{k}}+1) \left(\sqrt{2} \text{Hyp1}\left(\frac{1}{4}
   \left(2-\sqrt{2}\right)\right)^2 J+1\right)+2 J S\right)^2}\\
x_s^{\mathrm{II/III}}&=&-\left( \frac{JS}{2}+\frac{1}{6}\right):\nonumber \\
C_s^{\mathrm{II/III}}&=&-\frac{(324-324 i) \sqrt{3} \text{Hyp1}\left(\frac{1}{4} \left(2+5 i \sqrt{2}\right)\right)
   J^2 S \left(9 i \text{Hyp1}'\left(\frac{1}{4} \left(2+5 i
   \sqrt{2}\right)\right)+\sqrt{2} \text{Hyp1}\left(\frac{1}{4} \left(2+5 i
   \sqrt{2}\right)\right)\right)}{\left(6 J S+(6 \epsilon_{\mathbf{k}}+1) \left(1-3 i \sqrt{2}
   \text{Hyp1}\left(\frac{1}{4} \left(2+5 i \sqrt{2}\right)\right)^2J\right)\right)^2}   \\
x_s^{\mathrm{III/IV}}&=&-\left( \frac{JS}{2}-\frac{1}{6}\right):\nonumber \\
C_s^{\mathrm{III/IV}}&=&\frac{(324+324 i) \sqrt{3} \text{Hyp1}\left(\frac{1}{4} \left(2-5 i \sqrt{2}\right)\right)
   J^2 S \left(\sqrt{2} \text{Hyp1}\left(\frac{1}{4} \left(2-5 i \sqrt{2}\right)\right)-9 i
   \text{Hyp1}'\left(\frac{1}{4} \left(2-5 i \sqrt{2}\right)\right)\right)}{\left(6 J S+(6
   \epsilon_{\mathbf{k}}-1) \left(1-3 i \sqrt{2} \text{Hyp1}\left(\frac{1}{4} \left(2-5 i
   \sqrt{2}\right)\right)^2 J\right)\right)^2}  \\
x_s^{\mathrm{IV/I}}&=&-\left( \frac{JS}{2}-\frac{1}{2}\right):\nonumber \\
C_s^{\mathrm{IV/I}}&=&\frac{4 \sqrt[4]{-1} \text{Hyp1}\left(\frac{1}{4} \left(2-\sqrt{2}\right)\right) J^2 S
   \left(3 \sqrt{2} \text{Hyp1}'\left(\frac{1}{4} \left(2-\sqrt{2}\right)\right)+2
   \text{Hyp1}\left(\frac{1}{4} \left(2-\sqrt{2}\right)\right)\right)}{\left((2
   \epsilon_{\mathbf{k}}-1) \left(\sqrt{2} \text{Hyp1}\left(\frac{1}{4}
   \left(2-\sqrt{2}\right)\right)^2 J-1\right)-2 J S\right)^2}
\end{eqnarray}
\end{widetext}

\end{document}